\documentclass[12pt]{article}

\usepackage{graphicx}
\usepackage{comment}
\usepackage{color}
\usepackage{slashed}
\usepackage{bm}
\usepackage[pdfstartview=FitH,colorlinks=true,linkcolor=black,anchorcolor=black,citecolor=black,urlcolor=black]{hyperref}
\usepackage{amsmath,amssymb,titling,authblk}
\usepackage{slashed}
\usepackage{dsfont}
\usepackage{amscd}
\usepackage[normalem]{ulem}
\usepackage{appendix}
\usepackage{bbold}
\usepackage{lipsum}
\usepackage{amsthm}
\usepackage{lineno}
\usepackage{pstricks}
\usepackage[utf8]{inputenc}
\usepackage[top=2.3cm,right=2.3cm,left=2.3cm,bottom=2.3cm]{geometry}
\usepackage[numbers,sort&compress]{natbib}
\definecolor{verde}{cmyk}{.83,.21,1,.08}
\definecolor{darkorchid}{rgb}{0.6, 0.2, 0.8}
\definecolor{darkgreen}{rgb}{0,.5,0}

\def\({\left(}
\def\){\right)}
\def\[{\left[}
\def\]{\right]}

\newcommand{\be}{\begin{equation}}
	\newcommand{\ee}{\end{equation}}
\newcommand{\bea}{\begin{eqnarray}}
	\newcommand{\eea}{\end{eqnarray}}

\begin{document}
	
	\title{Advances in kappa deformed electrodynamics}
	
	\author[1]{O. Abla}
	\author[2]{M. J. Neves}
	\affil[ ]{}
	\affil[1]{\textit{\footnotesize Departamento de Física, Centro Federal de Educação Tecnológica de Minas Gerais, Av. Amazonas 7675, CEP 30510-000, Belo Horizonte, MG, 
			Brazil.}}
	\affil[2]{\textit{\footnotesize Departamento de Física, Universidade Federal Rural do Rio de Janeiro, BR 465-07, CEP 23890-971, Seropédica, RJ, Brazil.}}
	\affil[ ]{}
	\affil[ ]{\footnotesize e-mail: \texttt{olavoabla@cefetmg.br, mariojr@ufrrj.br}}
	
	\maketitle
	
	\begin{abstract}
			The so called $\kappa$-Minkowski spacetime is a prominent candidate for a quantum gravity space, and its properties have been the subject of investigation over the last thirty years. In this work, we propose classical electrodynamics equations that satisfy the dispersion relation derived from the bicrossproduct approach to the $\kappa$-Poincaré algebra.
			We explore the conservation laws, a new gauge symmetry, and the solutions for the scalar and vector potentials under the influence of the deformed parameter. The characteristics of the vacuum as a material medium are shown in terms of the refractive index and the group and phase velocities. The Green's functions associated with the wave operator in the bicrossproduct basis are calculated in both the retarded and advanced prescriptions. Finally, we discuss how the deformation scale affects the causality of the theory.
	\end{abstract}
\section{Introduction}
\label{sec1}

Perspectives on quantum gravity theories suggest that the smooth and continuous structure of spacetime becomes smeared at the Planck scale, implying a transition to a non-local nature \cite{DoplicherCMP}. One may consider these effects as being induced by an external field, where the spacetime observables are governed by noncommutative (NC) relations \cite{SzaboReview}
\begin{equation}
	[\, \hat{x}^{\mu}\, , \, \hat{x}^{\nu}\,]=i\,\Theta^{\mu\nu}(x) \; ,
\end{equation}
where $\Theta^{\mu\nu}(x)=-\Theta^{\nu\mu}(x)$ is an antisymmetric tensor, that in general depends on the spacetime coordinates. These approaches to quantum gravity \cite{seibergwitten99,Rivelles03,szabo2006,su(2)star3,Kontsevich} have motivated investigations of NC geometry in other field-theoretical models \cite{Nekrasov,ChaichianQM,Carroll,Neves1,Jabbari,Jonke1,PhysRep,PRLtwist,kappastar1,NCED3,twistkappa,FabianoJHEP}, equivalent to deformation quantization formalism, when one considers the star product of smooth functions $f(x)$ and $g(x)$ on a Poisson manifold \cite{Kontsevich}
\begin{equation}\label{Kontsevich}
	f \star g:=f \,\cdot\, g+\frac{i}{2}\,\Theta^{\mu\nu}(x) \, \partial_{\mu}f \, \partial_{\nu}g \, + \, \mathcal{O}(\Theta^2) \; ,
\end{equation}
in which the Poisson bracket is defined by
\begin{equation}\label{PB}
	\{\,f\,,\,g\,\}=\Theta^{\mu\nu}(x) \, \partial_{\mu}f \, \partial_{\nu}g \; ,
\end{equation}
where $\partial_{\mu}=\partial/\partial x^{\mu}$, and $\Theta$ is the antisymmetric Poisson bivector. The classical limit of the star product, {\it i.e.}, $\Theta\rightarrow0$, yields the standard pointwise product between functions, while the semiclassical limit is given by
\begin{equation}\label{sclimit}
	\{ \, f \, , \, g \, \}=\lim_{\Theta\to0} \, \frac{[ \, f \, , \, g \, ]_\star}{i\Theta} \; ,
\end{equation}
which inspired the formulation of {\it Poisson gauge theory} \cite{KKV1,KKV3,Kupriyanov:2021aet,Sharapov,Vlad2026,AN25,Kurkov2026,Kup2024} (see \cite{Kurkovreview} for a review), also known as {\it Poisson electrodynamics}, which represents the semi-classical limit of full U$(1)$ NC gauge theory. Several studies have used this nomenclature within the context of SU($\infty$) Yang-Mills theories \cite{SU(infty)2,SU(infty)3}. This framework is isomorphic to the infinite-dimensional Lie algebra of symplectic diffeomorphisms on the $S^2$-sphere. Consequently, the gauge potentials can be parameterized by the surface of an internal sphere at every spacetime point. This represents the symmetry of the canonically quantized relativistic Dirac membrane in the light-cone gauge after gauge fixing \cite{Supermembrane,Supermembrane2}. At that time, this framework was not directly related to NC gauge theories, where it was used in the $2000$s \cite{SU(infty)4}, until Kupriyanov and collaborators established the modern meaning of such gauge theories.
One of the deformed structures of interest in quantum gravity is based on $\kappa$-Poincaré algebra, and one of the realizations used in NC models, for the $\kappa$-Minkowski spacetime is given by \cite{Jonke1}
\begin{equation}
	[\, \hat{x}^0\, , \, \hat{x}^i\,]=\frac{i}{\kappa} \, \hat{x}^i 
	\quad ,\quad 
	[\, \hat{x}^i\, , \, \hat{x}^j\,]=0 \, ,
\end{equation}
where $\kappa$ is a parameter with mass dimension, and the limit $\kappa \rightarrow \infty$ recovers the usual commutative spacetime. 
This structure of NC spacetime is one of ways to the understanding of a quantum gravity theory \cite{Zakrzewski,Pachol}.   
Another possibility of introducing a mass scale (or length) scale in the spacetime structure is through of the Poincaré algebra. In this context, 
the Poincaré algebra is so deformed by a $q$-parameter with length dimension that is related to $\kappa$-parameter by $q=(2\kappa)^{-1}$. This 
deformation is understood as part of a rich mathematical structure called {\it Hopf algebra}, also known as quantum group, see the refs. 
\cite{ChaichianDemishev96,Majid95}. Hopf algebra is a mathematical framework used in the description of generalized symmetries beyond the 
Lie algebras \cite{ChaichianDemishev96}. Lukierski, Ruegg, Nowicki
and Tolstoy obtained a deformation of the Poincaré algebra that is known as kappa-deformed Poincaré algebra, from the $\kappa$ used to denote the
deformation parameter with dimension of mass \cite{LukierskiRueggNowickiTolstoy91,LukierskiNowickiRuegg92}. 
This is the simplest deformation of the Poincaré algebra which gives rise to
a quantum group in which it provides a theoretical laboratory for QFT investigations \cite{MajidRuegg94}.
The usual Poincaré algebra and all the known spacetime symmetries are so reobtained in the limit of $q\rightarrow 0$, or $\kappa \rightarrow \infty$. 
The algebraic sector of the $\kappa$-deformed Poincaré algebra is defined by commutation relations
between generators of time $(P^{0})$ and space $({\bf P})$ translations, as well as, the generators of
space rotations and boosts. The first kappa-deformation in the literature is known as
{\it standard basis dispersion relation} that has the Casimir invariant \cite{LukierskiRueggNowickiTolstoy91,LukierskiNowickiRuegg92}
\begin{equation}\label{RelDispBasep}
	{\cal C}_{\kappa}={\bf P}^2-(2\kappa)^{2}\sinh^2\left(\frac{P_0}{2\kappa}\right) \; ,
\end{equation}
where ${\cal C}_{\kappa}$ is element of the $\kappa$-Poincaré algebra that commutes with all the generators. 
If ${\cal C}_{\kappa}$ is proportional to the identity operator, {\it i.e.}, ${\cal C}_{\kappa}=-m^2$, the 
Casimir operator leads to dispersion relation for a massive particle $(m^2)$ in the standard basis
\begin{equation}\label{RelDispBasepm}
	(2\kappa)^{2}\sinh^2\left(\frac{P_0}{2\kappa}\right)-{\bf P}^2=m^2 \; .
\end{equation}
Using in (\ref{RelDispBasepm}) the usual representations $P^0\mapsto i\, \partial_t$ and
${\bf P}\mapsto -\,i \, \nabla$, we obtain the kappa-deformed Klein-Gordon operator in the standard basis. Such choice of deformation 
inspired investigations asking about how the length $q$-parameter could affect the causality in light signals \cite{Neves1}.

Other possibility of investigations is on kappa-deformation of Poincaré algebra known as the {\it bicrossproduct basis dispersion relation}, 
or {\it Majid-Ruegg product} \cite{MajidRuegg94,Majid95}. The main motivation for the study of this quantum group is related to it unification 
with the NC spacetime through a star product that allows the construction of an action for scalar field relativistically invariant under the 
$\kappa$-Poincaré algebra \cite{Meljanac}.     
In this paper, we explore the classical electrodynamics based on the kappa-deformation in the bicrossproduct basis. Following the recent ref. \cite{Vlad2026}, 
the Casimir invariant in this basis is given by
\begin{equation}
	{\cal C}_{\kappa}={\bf P}^2 \, e^{P^{0}/\kappa}-(2\kappa)^{2}\sinh^2\left(\frac{P_0}{2\kappa}\right) \; ,
\end{equation}
in which using the last substitutions for the generators $(P^{0},{\bf P})$, and ${\cal C}_{\kappa}=-m^2$, 
the Klein-Gordon operator and the Dirac equation in the bicrossproduct basis has been discussed in the ref. \cite{Vlad2026}. 
The correspondent Casimir operator on-shell mass in the bicrossproduct basis is
\begin{equation}\label{KGop}
	(2\kappa)^{2}\sinh^2\left(\frac{P_0}{2\kappa}\right)-{\bf P}^2 \, e^{P^{0}/\kappa}=m^2 \; ,
\end{equation}
The plane wave solutions for the KG operator (\ref{KGop}) leads to the dispersion relation in the
bicrossproduct basis
\begin{equation}\label{reldispbp2}
	\left[\,\frac{1}{q}\sinh(q\omega)\,\right]^{2}-e^{2q\omega}\,{\bf k}^2=m^2 \; ,
\end{equation}
for any $\omega$-frequency and ${\bf k}$-wave vector, with $q=(2\kappa)^{-1}$. We investigate the effects of a classical electrodynamics based on the dispersion 
relation (\ref{reldispbp2}), when $m=0$. Thereby, we propose the $\kappa$-deformed equations (in bicrossproduct basis) 
for the electric and magnetic field that satisfy the massless dispersion relation (\ref{KGop}) through the plane wave solutions.                  
The paper is organized as follows: In the Section \ref{sec2}, we introduce the $\kappa$-deformed electrodynamics. In Section \ref{sec3}, we obtain the relations between fields and potentials, a $\kappa$-deformed gauge symmetry, and the machinery for compute the retarded and advanced Green functions. In Section \ref{sec4}, we show the $\kappa$-deformed structure of the vacuum as material medium, as the dispersion relations, refractive index, and group/phase velocities. In Section \ref{sec5}, we calculate the $\kappa$-deformed Green functions. For end, in Section \ref{sec6}, we present the conclusions.
We use the natural units $\hbar=c=1$ throughout this work. Greek letters correspond to $\mu,\nu=\{ \, 0,1,2,3 \, \}$, where the metric tensor in has the signature  $\eta^{\mu\nu}=(+1,-1,-1,-1)$.

\section{The kappa deformed electrodynamics}
\label{sec2}
Following using the same mass-shell condition (\ref{reldispbp2}), exactly the same for $m=0$ in Ref. \cite{Vlad2026}, the equations for kappa-deformed electrodynamics that satisfy such condition are given by
\begin{subequations}
	\begin{eqnarray}
		\nabla\cdot ( e^{i\,2q\,\partial_{t}} \, {\bf E})=\rho \; ,
		\label{GaussLaw}
		\\
		\nabla\cdot ( e^{i\,2q\,\partial_{t}} \, {\bf B})=0 \; ,
		\label{GaussLawB}
		\\
		\nabla\times (e^{i\,q\,\partial_{t}} \, {\bf E})+\partial_{q}{\bf B}={\bf 0} \; ,
		\label{FLL}
		\\
		\nabla\times (e^{i\,q\,\partial_{t}} \, {\bf B})={\bf J}+\partial_{q}{\bf E} \; , 
		\label{AML}
	\end{eqnarray}
\end{subequations}
where $\rho$ and ${\bf J}$ are the sources, ${\bf E}$ and ${\bf B}$ are the electric and magnetic fields, and $\partial_{q}$ is the time-differential 
operator set by the infinity series
\begin{equation}
	\partial_{q}=\frac{1}{q}\,\sin(q\partial_{t})=\sum_{n=0}^{\infty}\frac{(-1)^{n} \, q^{2n}}{(2n+1)!}\,\partial_{t}^{2n+1} \; .
\end{equation}
In the limit $q \rightarrow 0$, the usual Maxwell electrodynamics is recovered, with $\lim_{q\rightarrow 0}\partial_{q}=\partial_{t}$. When the fields are static, the set of equations reduce to the usual electrostatic and magnetostatic frameworks.
Since these are complex field equations, the vector fields and the sources are considered complex functions. The sources $\rho$ and ${\bf J}$ have real and imaginary parts that contain 8 degrees of freedom. For the 
fields ${\bf E}$ and ${\bf B}$, the 6 components imply 12 degrees of freedom for their real and imaginary parts. In general, we have a total of 20 degrees of freedom for both the fields and sources. We can check that the current and charge densities satisfy the 
continuity equation
\begin{equation}\label{EqJ}
e^{i\,2q\,\partial_{t}}(\nabla\cdot{\bf J})+\partial_{q}\rho=0 \; .
\end{equation}
The static limit in (\ref{EqJ}) leads to stationary current condition $\nabla\cdot{\bf J}=0$. 
For the vacuum case, with $\rho=0$ and ${\bf J}={\bf 0}$, we take the rotational operation on the Faraday-Lenz law (\ref{FLL}) by the operator $\nabla \, e^{i\,q\,\partial_{t}}$, and substituting (\ref{GaussLaw}) and (\ref{AML}), we obtain the electric wave equation
\begin{equation}\label{EqwaveE}
	\left[ \, \frac{1}{q^2} \sin^2\left(q\,\partial_{t}\right)-e^{i\,2q\,\partial_{t}}\,\nabla^{2} \, \right]{\bf E}={\bf 0} \; .
\end{equation}
Similarly, using the same operation of $\nabla \, e^{i\,q\,\partial_{t}} \times$ on the Ampère-Maxwell law (\ref{AML}), with the help of (\ref{GaussLawB}) and (\ref{FLL}), the magnetic field satisfies the wave equation
\begin{equation}\label{EqwaveB}
	\left[ \, \frac{1}{q^2} \sin^2\left(q\,\partial_{t}\right)-e^{i\,2q\,\partial_{t}}\,\nabla^{2} \, \right]{\bf B}={\bf 0} \; .
\end{equation}  
Substituting the plane wave solutions 
\begin{equation}\label{planewavesol}
	{\bf E}={\bf E}_{0} \, e^{i({\bf k}\cdot{\bf r}-\omega t)}
	\quad , \quad
	{\bf B}={\bf B}_{0} \, e^{i({\bf k}\cdot{\bf r}-\omega t)} \; ,
\end{equation}
in the Eqs. (\ref{EqwaveE}) and $(\ref{EqwaveB})$, for the constant amplitudes ${\bf E}_{0}$ and ${\bf B}_{0}$, 
we obtain the dispersion relation
\begin{equation}\label{reldisp}
	\frac{1}{q^2} \sinh^2\left( q\,\omega \right) - e^{2q\,\omega}\,{\bf k}^2=0 \; ,
\end{equation}
where $\omega$ is the frequency, and ${\bf k}$ is the wave vector. Thus, we can affirm that the Eqs. (\ref{GaussLaw})-(\ref{AML}) satisfy 
dispersion relations in the bicrossproduct basis for the massless case. The equations (\ref{GaussLaw})-(\ref{AML}) in the frequency space (in vacuum) are given by
\begin{subequations}
\begin{eqnarray}
{\bf k}\cdot{\bf E}_{0} \!\!&=&\!\! {\bf k}\cdot{\bf B}_{0} = 0 \; ,
\label{kEkB}
\\
{\bf k}\times{\bf E}_{0} \!\!&=&\!\! e^{-q\,\omega} \, \frac{1}{q}\sinh\left( q\,\omega \right){\bf B}_{0} \; ,
\label{BkE}
\\
{\bf k}\times{\bf B}_{0} \!\!&=&\!\! -\,e^{-q\,\omega} \, \frac{1}{q}\sinh\left( q\, \omega \right){\bf E}_{0} \; ,
\label{EkB}
\end{eqnarray}
\end{subequations}
and show that the vectors ${\bf k}$, ${\bf E}_{0}$ and ${\bf B}_{0}$ remain perpendicular to one another, and as a consequence, the $q$-deformation does not affect the transversality of the electromagnetic plane wave. Notice that the $q$-deformation appears coupled to the time derivatives in the classical field equations. The original frequency $\omega$ of the model is related to the generator of translations $(P_{0})$ in the $\kappa$-Poincar\'e algebra through the correspondence $P_{0} \mapsto i\,\partial_{t} \rightarrow \omega$ for the plane wave solutions (\ref{planewavesol}). The set of equations (\ref{kEkB})-(\ref{EkB}) shows 
	a dynamical equivalence in the momentum space in relation to non-deformed theory if we define the $q$-frequency as $\omega_{q}=e^{-q\,\omega} q^{-1} \sinh\left( q\, \omega \right)$. This dynamical equivalence does not imply physical equivalence since the observables are different comparing the $q$-deformed with the usual theory.
	For the plane wave solutions, the relations (\ref{kEkB})-(\ref{EkB}) keep the same counting of the degrees of freedom for the fields ${\bf E}_{0}$ and ${\bf B}_{0}$ from the usual Maxwell electrodynamics. Since that in (\ref{kEkB})-(\ref{EkB}) we have four independent equations that lead to two complex polarizations, and consequently, 
	we have four real degrees of freedom. 


%
%

%
Considering ${\bf E}$ and ${\bf B}$ complex fields in the equations (\ref{GaussLaw})-(\ref{AML}), we write the stationary solutions 
${\bf E}={\bf E}_{0}({\bf r})\,e^{-i\omega t}$ and ${\bf B}={\bf B}_{0}({\bf r})\,e^{-i\omega t}$, where ${\bf E}_{0}({\bf r})$ and ${\bf B}_{0}({\bf r})$
are fields that depend only on the spatial coordinates, as well as, the sources $\rho=\rho_{0}({\bf r})\,e^{-i\omega t}$ and ${\bf J}={\bf J}_{0}({\bf r})\,e^{-i\omega t}$, 
all with the same $\omega$-frequency. Under these solutions, the equations (\ref{FLL}) and (\ref{AML}) are, respectively, given by   
\begin{subequations}
	\begin{eqnarray}
		\nabla\times{\bf E}_{0}-e^{-q\,\omega} \, \frac{i}{q} \, \sinh(q\omega) \, {\bf B}_{0}({\bf r}) \!\!&=&\!\! {\bf 0} \; ,
		\label{FLLSt}
		\\
		\nabla\times {\bf B}_{0}+e^{-q\,\omega} \, \frac{i}{q} \, \sinh(q\omega) \, {\bf E}_{0}({\bf r}) \!\!&=&\!\! e^{-q\,\omega} \, {\bf J}_{0}({\bf r}) \; . 
		\label{AMLSt}
	\end{eqnarray}
\end{subequations}

Since the spatial part is unaltered, the complex Poynting vector is ${\bf S}=({\bf E}_{0}\times {\bf B}_{0}^{\ast})/2$. The Poynting theorem can be obtained multiplying (\ref{FLLSt}) with the scalar product of ${\bf B}_{0}^{\ast}$ on both sides, and the complex conjugate of (\ref{AMLSt}) with the scalar product of ${\bf E}_{0}$, we have
\begin{subequations}
	\begin{eqnarray}
		{\bf B}_{0}^{\ast}\cdot(\nabla\times{\bf E}_{0})-e^{-q\,\omega} \, \frac{i}{q} \, \sinh(q\omega) \, 
		|{\bf B}_{0}({\bf r})|^2 \!\!&=&\!\! 0 \; ,
		\label{Rel1}
		\\
		{\bf E}_{0}\cdot(\nabla\times{\bf B}_{0})-e^{-q\,\omega} \, \frac{i}{q} \, \sinh(q\omega) \, |{\bf E}_{0}({\bf r})|^2
		\!\!&=&\!\! e^{-q\,\omega} \, {\bf J}_{0}^{\ast}({\bf r})\cdot{\bf E}_{0}({\bf r}) \; .
		\label{Rel2}
	\end{eqnarray}
\end{subequations}

Using the identity
\begin{equation}
	\nabla\cdot {\bf S}=\frac{1}{2} \left[ \, {\bf B}_{0}^{\ast}\cdot(\nabla\times{\bf E}_{0})-{\bf E}_{0}\cdot(\nabla\times{\bf B}^{\ast}_{0}) \, \right] \; ,
\end{equation}
and substituting the expressions (\ref{Rel1}) and (\ref{Rel2}), we obtain the Poynting theorem 
\begin{equation}
	\nabla\cdot {\bf S}+e^{-q\,\omega} \, \frac{i}{2q} \, \sinh(q\omega) \, \left( \, |{\bf B}_{0}({\bf r})|^2 - |{\bf E}_{0}({\bf r})|^2 \, \right)= -\frac{1}{2} \, e^{-q\,\omega} \, {\bf J}_{0}^{\ast}({\bf r})\cdot{\bf E}_{0}({\bf r}) \; .
\end{equation}

In the integral form is
\begin{equation}\label{fluxS}
	\oint_{{\cal S}} {\bf S}\cdot\hat{{\bf n}} \, dA=-\frac{1}{2} \, e^{-q\,\omega} \int_{{\cal R}} d^{3}{\bf r} \, {\bf J}_{0}^{\ast}({\bf r})\cdot{\bf E}_{0}({\bf r})- \frac{i}{q} \, (1-e^{-2q\omega}) \int_{{\cal R}} d^{3}{\bf r} \, \left[ \, \frac{1}{4} \, |{\bf B}_{0}({\bf r})|^2 - \frac{1}{4} \, |{\bf E}_{0}({\bf r})|^2 \, \right] \; .
\end{equation}

The real part of the flux of ${\bf S}$ is interpreted as the power that goes out from the region ${\cal R}$, and the imaginary part of (\ref{fluxS}) is the reactive power that depends on the difference between magnetic and electric energies. For a small $q$-parameter, the contribution of the imaginary part goes linearly with the $\omega$-frequency, as we expect. For $q$-fixed and high $\omega$-frequency, the reactive power approaches the limit of $q^{-1}$ times the difference between the energies.     
\section{Potentials and gauge symmetry}
\label{sec3}
From the equation (\ref{GaussLawB}), we define the relation between the magnetic field and vector potential ${\bf A}$
\begin{equation}\label{campoB}
	{\bf B}=e^{-i\,2q\,\partial_{t}}(\nabla \times {\bf A}) \; ,
\end{equation}
and substituting in the Faraday-Neumann-Lenz law, the electric field is given by scalar ($V$) and vector potentials
\begin{equation}\label{campoE}
	{\bf E}=-e^{-i\,q\,\partial_{t}}\,\nabla V - e^{-i\,3q\,\partial_{t}} \, \partial_{q}{\bf A} \; . 
\end{equation}

Substituting (\ref{campoE}) in Eq. (\ref{GaussLaw}), and (\ref{campoB}) in Eq. (\ref{AML}), we obtain the equations for the potentials
\begin{subequations}
	\begin{eqnarray}
		e^{i\,q\,\partial_{t}}\,\nabla^2V+e^{-i\,q\,\partial_{t}} \, \partial_{q}(\nabla\cdot{\bf A}) \!&=&\! -\rho \; ,
		\label{EqV}
		\\
		e^{-i\,q\,\partial_{t}}\,\nabla\left(\nabla\cdot{\bf A}+\partial_{q}V\right)-e^{-i\,q\,\partial_{t}}\,\nabla^2{\bf A}+e^{-i3\,q\,\partial_{t}}\,\partial_{q}^2{\bf A} \!&=&\! {\bf J} \; .
		\label{EqA}
	\end{eqnarray}
\end{subequations}

The fields ${\bf E}$ and ${\bf B}$ are gauge invariant under the $\kappa$-transformations
\begin{equation}\label{transfgaugeVA}
	V \, \longmapsto \, V^{\prime}=V-e^{-i\,2q\,\partial_{t}}\,\partial_{q}f\,,\quad
	{\bf A} \, \longmapsto \, {\bf A}^{\prime}={\bf A}+\nabla f \; ,
\end{equation}
for an arbitrary function $f$ of the spacetime coordinates. Immediately, we have the gauge invariance ${\bf E}^{\prime}={\bf E}$ 
and ${\bf B}^{\prime}={\bf B}$, that is preserved in the Eqs. (\ref{GaussLaw})-(\ref{AML}). It is important to highlight that the Eqs. (\ref{GaussLaw})-(\ref{AML}) set an effective model in which the propagation of the electromagnetic fields can be investigated. The construction of an action associated with these equations (\ref{GaussLaw})-(\ref{AML}), that is invariant under all the symmetries of the $\kappa$-Poincar\'e algebra, is a pertinent challenge in the study of the Abelian gauge theory within the framework of quantum groups. The gauge transformations (\ref{transfgaugeVA}) presented here leave the classical field equations (\ref{GaussLaw})-(\ref{AML}) invariant. Using the gauge transformations, we can 
check that the combination $\nabla\cdot{\bf A}+\partial_{q}V$ satisfies the relation
\begin{equation}
	\nabla\cdot{\bf A}^{\prime}+\partial_{q}V^{\prime}=\nabla\cdot{\bf A}+\partial_{q}V +e^{-i\,2q\,\partial_{t}}\left( \, e^{i\,2q\,\partial_{t}}\,\nabla^2f-\partial_{q}^2f \, \right) \; .
\end{equation}

Fixing the {\it $\kappa$-Lorenz gauge condition} $\nabla\cdot{\bf A}^{\prime}+\partial_{q}V^{\prime}=\nabla\cdot{\bf A}+\partial_{q}V=0$, the $f$-function satisfies the same free wave equations from (\ref{EqwaveE}) and (\ref{EqwaveB}). Thus, choosing the $\kappa$-Lorenz gauge $\nabla\cdot{\bf A}+\partial_{q}V=0$, the Eqs. (\ref{EqV}) and (\ref{EqA}) are simplified as
\begin{subequations}
	\begin{eqnarray}
		\left[ \, \frac{1}{q^2} \sin^2\left(q\,\partial_{t}\right)-e^{i\,2q\,\partial_{t}}\,\nabla^{2} \, \right] V \!&=&\! e^{i\partial_{t}/(2\kappa)}\,\rho \; ,
		\label{EqVLorenz}
		\\
		\left[ \, \frac{1}{q^2} \sin^2\left(q\,\partial_{t}\right)-e^{i\,2q\,\partial_{t}}\,\nabla^{2} \, \right] {\bf A} \!&=&\! e^{i3\partial_{t}/(2\kappa)}\, {\bf J} \; .
		\label{EqALorenz}
	\end{eqnarray}
\end{subequations}

The solutions for the Eqs. (\ref{EqVLorenz}) and (\ref{EqALorenz}) are given by
\begin{subequations}
	\begin{eqnarray}
		V({\bf x},t) \!\!&=&\!\! -\int d^3{\bf r}^{\prime} \int_{-\infty}^{\infty}dt^{\prime} \, G(t-t^{\prime},{\bf x}-{\bf x}^{\prime}) \, e^{i\,q\,\partial_{t^{\prime}}}\,\rho({\bf x}^{\prime},t^{\prime}) \; , 
		\label{solpotV}
		\\
		{\bf A}({\bf x},t) \!\!&=&\!\! -\int d^3{\bf r}^{\prime} \int_{-\infty}^{\infty}dt^{\prime} \, G(t-t^{\prime},{\bf x}-{\bf x}^{\prime}) \, e^{i \, 3q\,\partial_{t^{\prime}}}\,{\bf J}({\bf x}^{\prime},t^{\prime}) \; ,
		\label{solpotA}
	\end{eqnarray}
\end{subequations}
where the Green function $G(t-t^{\prime},{\bf x}-{\bf x}^{\prime})$ satisfies the Green equation
\begin{equation}\label{equacaogreenbb}
	\left[\, \frac{1}{q^2}\,\sin^2\left(q\,\partial_{t}\right)
	-e^{i\,2q\,\partial_{t}} \,\nabla^{2} \, \right]
	G(t-t^{\prime},{\bf x}-{\bf x}^{\prime})=
	-\,\delta^3({\bf x}-{\bf x}^{\prime})\, \delta(t-t^{\prime}) \; .
\end{equation} 

The solutions of the Eq. (\ref{equacaogreenbb}) for the Green function $G$ allow us to obtain solutions for the scalar 
and vector potentials ruled by (\ref{solpotV}) and (\ref{solpotA}), respectively. We will do it in the Section \ref{sec5}.

\section{Vacuum as a dispersive medium}
\label{sec4}
Returning to dispersion relation (\ref{reldisp}), the frequency solutions are given by
\begin{equation}\label{frequenciasbbck<}
	\omega^{(\pm)}({\bf k})=-\frac{1}{2q}\ln\left(1 \mp 2q|{\bf k}| \right) \; ,
\end{equation}
that is valid for the condition $|{\bf k}|<(2q)^{-1}$. For the condition $|{\bf k}|>(2q)^{-1}$, just the positive frequency $\omega^{(+)}$ is modified as
\begin{equation}\label{frequenciasbbck>}
	\omega^{(>)}({\bf k})=-\frac{1}{2q}\ln\left(2q|{\bf k}|-1\right)-\frac{i\pi}{2q} \; .
\end{equation}

When $|{\bf k}|=(2q)^{-1}$, the frequency on the momentum cutoff is $\omega=-(2q)^{-1}\ln(2)$. The frequencies (\ref{frequenciasbbck<}) are asymmetric, and consequently, it exhibits a charge breaking symmetry between particles and antiparticles. The frequency $\omega^{(>)}$ assumes complex values with an imaginary part 
that depends on the $q$-parameter. 
Defining the refractive index of medium as $n=|{\bf k}|/\omega$, the positive solution of (\ref{frequenciasbbck<}) and (\ref{frequenciasbbck>})
yield the results :
\begin{subequations}
	\begin{eqnarray}
		n^{(+)}({\bf k}) \!\!&=&\!\! \frac{-2q\,|{\bf k}|}{\ln(1-2q\,|{\bf k}|)} \; ,
		\label{nplus}
		\\
		n^{(>)}({\bf k}) \!\!&=&\!\! \frac{-\,2q\,|{\bf k}|}{\ln(2q\,|{\bf k}|-1)+i\,\pi}=-\,2q\,|{\bf k}| \left[ \, \frac{\ln(2q\,|{\bf k}|-1)-i\,\pi}{\ln(2q\,|{\bf k}|-1)^2+\pi^2} \, \right] \; ,
		\label{nmaior}
	\end{eqnarray}
\end{subequations}
in which the solution $n^{(>)}({\bf k})$ shows a real and imaginary parts that are expected due to (\ref{frequenciasbbck>}). 
In the limit $q\rightarrow 0$, we recover $\lim_{q\rightarrow 0} n^{(+)}({\bf k})=1$ and $\lim_{q\rightarrow 0} n^{(>)}({\bf k})=0$. 
For a very small $q$-parameter, the first correction of (\ref{nmaior}) is pure imaginary, {\it i. e.}, $n^{(>)}({\bf k}) \simeq iq\,|{\bf k}|/\pi\simeq i\,2q/\lambda$ with wavelength $(\lambda)$. In this condition of small $q$-parameter, the electric and magnetic amplitudes are attenuated by 
\begin{equation}
	{\bf E}^{(>)}\simeq {\bf E}_{0}\,e^{-(2q/\lambda)\,\omega \, (\hat{{\bf k}}\cdot{\bf r})}\,e^{-i\omega\,t}
	\hspace{0.2cm} , \hspace{0.2cm}
	{\bf B}^{(>)}\simeq {\bf B}_{0}\,e^{-(2q/\lambda)\,\omega \, (\hat{{\bf k}}\cdot{\bf r})}\,e^{-i\omega\,t} \; .
\end{equation}

The group velocity is obtained taking the differential of the dispersion relation (\ref{reldisp}), that leads to 
\begin{equation}\label{vg}
v_{g}=\frac{d\omega}{d|{\bf k}|}=\frac{2q\,|{\bf k}|\,e^{2q\,\omega}}{\sinh(2q\,\omega)-2q^2\,|{\bf k}|^2 \, e^{2q\,\omega}}\; ,
\end{equation}
that is real and positive only if $\omega > -(\kappa/2) \,\ln(1-|{\bf k}|^2/\kappa^2)$, when $|{\bf k}| < \kappa$. Thus, we can use only the solution $\omega^{(+)}({\bf k})$ 
from (\ref{frequenciasbbck<}). Substituting $\omega^{(+)}({\bf k})$ in (\ref{vg}), we can verify that the group velocity is
\begin{eqnarray}
\left. v_{g} \right|_{\omega^{(+)}} = \frac{1}{1-2q\,|{\bf k}|} \; ,
\end{eqnarray}
when $0< |{\bf k}| < (2q)^{-1}$, and we have $v_{g}>1$ with superluminal propagation \cite{NCED3}. The phase velocity for the free wave in vacuum is
\begin{equation}\label{vf}
v_{ph}^{(>)}(k)=\frac{\omega^{(+)}({\bf k})}{|{\bf k}|}=-\frac{1}{2q|{\bf k}|}\ln\left(1 - 2q\,|{\bf k}| \right)  \; ,
\end{equation}  
valid for $|{\bf k}|<(2q)^{-1}$.
\begin{figure}[htbp]
	\centering
	\begin{minipage}[b]{0.32\textwidth}
		\centering
		\includegraphics[width=\textwidth]{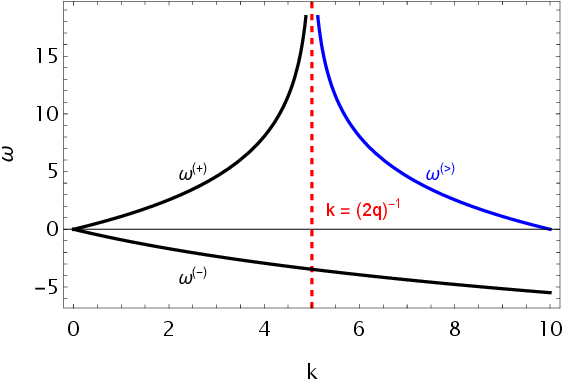}
	\end{minipage}
	\hfill
	\begin{minipage}[b]{0.32\textwidth}
		\centering
		\includegraphics[width=\textwidth]{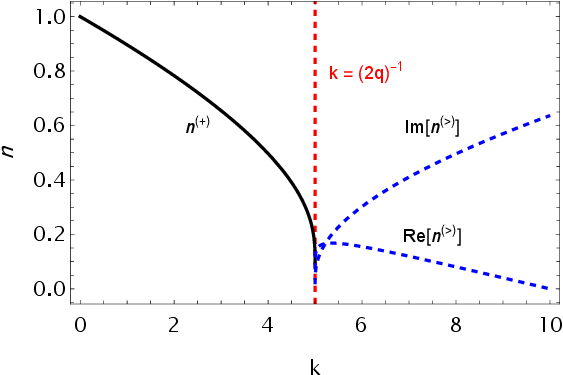}
	\end{minipage}
	\hfill
	\begin{minipage}[b]{0.32\textwidth}
		\centering
		\includegraphics[width=\textwidth]{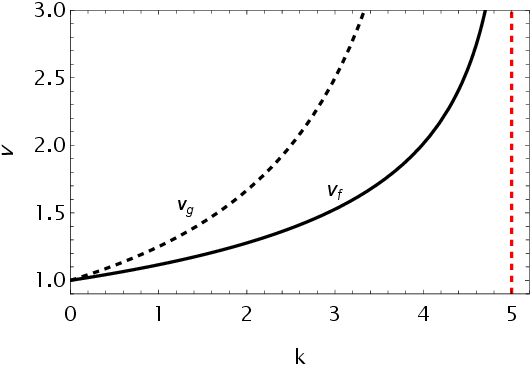}
	\end{minipage}
	
	\caption{Left panel: The $\omega$-frequencies versus $k$. Medium panel : The refractive index versus $k$. Right panel : Group and phase velocities versus $k$, for $k<(2q)^{-1}$. In all these plots, we use $q=0.1$ (in length unit). The dashed red line sets the cutoff momentum $k=(2q)^{-1}$. } 
	\label{fig1}
\end{figure}

The results of this section are plotted in the figure (\ref{fig1}). In the left panel, we show the frequencies $\omega^{(\pm)}({\bf k})$ (back lines) and 
$\omega^{(>)}({\bf k})$ (blue line) as functions of $k=|{\bf k}|$. The medium panel is the refractive index $n^{(+)}({\bf k})$ (black line), the real and imaginary 
parts (dashed blue lines) of $n^{(>)}({\bf k})$ as functions of $k$. The right panel sets the phase velocity as function of $k$. In all the images, we use
$q=0.1$ (in length units) for the $q$-parameter. The dashed red line means the cutoff momentum at $k=(2q)^{-1}$. The third figure shows how the phase velocity 
curve grows with $k$ in which the superluminal propagation is evident for $k < (2q)^{-1}$.

\section{Retarded and advanced Green functions}
\label{sec5}

In this section, we investigate the aspects of causality through the retarded and advanced Green functions that obey the equation (\ref{equacaogreenbb}).   
Using the Fourier transform, the Green function is written as the integrals 
\begin{equation}\label{FourierBDPC}
	G^{(\gamma)}(t-t^{\prime},{\bf x}-{\bf x}^{\prime})=\int\; \frac{d^3{\bf k}}{(2\pi)^3}
	\int_{(\gamma)} \frac{d\omega}{2\pi} \; \frac{e^{-ik\cdot(x-x^{\prime})}}{q^{-2}\sinh^{2}(q\omega)-e^{2q\omega}\,{\bf k}^2}  \; .
\end{equation}

For simplification of the computations, the integral (\ref{FourierBDPC}) can be rewritten into the form
\begin{equation}\label{FourierBDPCInt}
	G^{(\gamma)}(t-t^{\prime},{\bf x}-{\bf x}^{\prime})
	=\int\; \frac{d^3{\bf k}}{(2\pi)^3} \; e^{i{\bf k}\cdot({\bf x}-{\bf x}^{\prime})} \,
	{\cal G}^{(\gamma)}({\bf k},t-t^{\prime}) \; ,
\end{equation}
where we define the temporal part as
\begin{equation}\label{GIntomega}
	{\cal G}^{(\gamma)}({\bf k},t-t^{\prime})=\int_{(\gamma)}
	\frac{d\omega}{2\pi} \; \frac{e^{-i\omega  (t-t^{\prime})}}
	{q^{-2}\sinh^{2}(q\omega)-e^{2q\omega}\,{\bf k}^2} \; ,
\end{equation}
and $(\gamma)$ sets a contour in the $\omega$-complex plane. Notice that the denominator of
(\ref{GIntomega}) can be written in terms of exponential functions 
\begin{equation}\label{denominator}
	\frac{1}{q^{-2}\sinh^{2}(q\omega)-e^{2q\omega}\,{\bf k}^2}
	=\frac{4q^{2}e^{2q\omega}}{(1-4q^{2}\,{\bf k}^2)(e^{2q\omega})^{2}-2\,e^{2q\omega}+1} \; ,
\end{equation}
in which the term $1-4q^{2}\,{\bf k}^{2}$ assumes the values of $>0$, $<0$ and $=0$.
Consequently, the zeros in the denominator of (\ref{denominator}) yield the poles of the $\omega$-integration. 
When $1-4q^{2}\,{\bf k}^{2}=0$ the poles of the $\omega$-integration will have a momentum cutoff at
$|{\bf k}|=(2q)^{-1}$, as we mentioned in the previous section. In this case,
the poles on the momentum $|{\bf k}|=(2q)^{-1}$ are
\begin{equation}\label{polesk=kappa}
	\omega_{n}=-\frac{1}{2q}\ln(2)+\frac{i\pi n}{q} \; ,
\end{equation}
with with $n$ any integer number.
For the case of $1-4q^{2}\,{\bf k}^{2}>0$, {\it i.e.}, with
$|{\bf k}|<(2q)^{-1}$, the poles of the
$\omega$-integration are given by
\begin{equation}\label{polesk<kappa}
	\omega_{n}^{(\pm)}=\omega^{(\pm)}({\bf k})+\frac{i\pi n}{q} \; ,
\end{equation}
where $\omega^{(\pm)}$ are the asymmetric dispersion relations (\ref{frequenciasbbck<}). 
When $1-4q^{2}\,{\bf k}^{2}<0$, or $|{\bf k}|>(2q)^{-1}$, the negative poles of (\ref{polesk<kappa})
are unaltered, while the positive poles are given by
\begin{equation}\label{polesk>kappa}
	\omega_{n}^{(>)}=\omega^{(>)}({\bf k})-\frac{i\pi}{2q}+\frac{i\pi n}{q} \; ,
\end{equation}
where $\omega^{(>)}({\bf k})$ is the frequency (\ref{frequenciasbbck>}) above the momentum cutoff.
The choice in the Green function
(\ref{FourierBDPC}) of a contour $(\gamma)$ bypassing the poles at the points
$(\omega^{(\pm)}({\bf k}),\omega^{(>)}({\bf k}))$ on the $\omega$-real axis is equivalent
to displace them by infinitesimal imaginary quantities.
The displacement $(\omega^{(\pm)}({\bf k}),\omega^{(>)}({\bf k})) \mapsto (\omega^{(\pm)}({\bf k})-i\varepsilon,\omega^{(>)}({\bf k})-i\varepsilon)$ is interpreted as the retarded Green function, and the displacement $(\omega^{(\pm)}({\bf k}),\omega^{(>)}({\bf k}))\mapsto (\omega^{(\pm)}({\bf k})+i\varepsilon,\omega^{(>)}({\bf k})+i\varepsilon)$ is associated with the advanced Green function. For the causal Green function, the displacement is given by the {\it Feynman
	prescription}, in which $(\omega^{(\pm)}({\bf k}),\omega^{(>)}({\bf k})) \mapsto (\omega^{(\pm)}({\bf k})
\mp i\varepsilon,\omega^{(>)}({\bf k})-i\varepsilon)$. We denote $(+)$ the integration path that represents the $\kappa$-retarded Green function,
as well as, the symbol $(-)$ denotes the integration path associated with the $\kappa$-advanced
Green function. These integration paths are equivalent to substituting
$\omega \mapsto \omega+i\varepsilon$ for $\kappa$-retarded Green function, and
$\omega \mapsto \omega-i\varepsilon$ for $\kappa$-advanced Green function in the
$\omega$-integration of (\ref{GIntomega}). The Green functions can be calculated by
the usual complex variable technique of closing the contour $(\gamma)$ by a
semicircle on the upper or lower half plan, where the ratio of the semicircle goes to infinity. 
Firstly, we will calculate the $\kappa$-retarded
Green function, and by an analogous way, we will obtain the $\kappa$-advanced Green function.
%
%
%
\begin{figure}[!h]
	\begin{center}
		\newpsobject{showgrid}{psgrid}{subgriddiv=1,griddots=10,gridlabels=6pt}
		\begin{pspicture}(-2,-3)(4,3.3)
			\psset{arrowsize=0.2 2}
			%
			%
			\psline[linecolor=black,linewidth=0.4mm]{->}(0.5,-3.5)(0.5,3.4)
			\psline[linecolor=black,linewidth=0.4mm]{->}(-4,0)(5,0)
			%
			%
			\psline[linecolor=black,linewidth=0.5mm](-1.51,0)(-1.51,-0.15)
			%
			%
			\put(5,-0.3){$\Re(\omega)$}
			\put(0.7,3.1){$\Im(\omega)$}
			%
			%
			\put(-3.6,-0.7){$-(2q)^{-1}\ln(2)-i\varepsilon$}
			%
			%
			\put(-1.5,2.85){\circle*{0.1}}
			\put(-1.5,1.85){\circle*{0.1}}
			\put(-1.5,0.85){\circle*{0.1}}
			\put(-1.5,-0.15){\circle*{0.1}}
			\put(-1.5,-1.15){\circle*{0.1}}
			\put(-1.5,-2.15){\circle*{0.1}}
			\put(-1.5,-3.15){\circle*{0.1}}
			%
			%
			%
		\end{pspicture}
		\vspace{0.4cm}
		\caption{Poles in the $\omega$-complex plane for the $\kappa$-retarded
			Green function on the cutoff momentum $|{\bf k}|=(2q)^{-1}$.}
	\end{center}
\end{figure}
%
%
%
\begin{figure}[!h]
	\begin{center}
		\newpsobject{showgrid}{psgrid}{subgriddiv=1,griddots=10,gridlabels=6pt}
		\begin{pspicture}(-2,-3)(4,4.5)
			\psset{arrowsize=0.2 2}
			%
			%
			\psline[linecolor=black,linewidth=0.4mm]{->}(0.5,-3.5)(0.5,3.3)
			\psline[linecolor=black,linewidth=0.4mm]{->}(-4,0)(5,0)
			%
			%
			\psline[linecolor=black,linewidth=0.5mm](-1.5,0)(-1.5,-0.15)
			\psline[linecolor=black,linewidth=0.5mm](2,0)(2,-0.15)
			%
			%
			\put(4.5,-0.5){$\Re(\omega)$}
			\put(0.7,3){$\Im(\omega)$}
			\put(3,1.5){$|{\bf k}|<(2q)^{-1}$}
			\put(2.2,-0.5){$\omega^{(+)}({\bf k})-i\varepsilon$}
			\put(-3.6,-0.5){$\omega^{(-)}({\bf k})-i\varepsilon$}
			%
			\put(2,3){\circle*{0.1}}
			\put(2,2){\circle*{0.1}}
			\put(2,1){\circle*{0.1}}
			\put(2,-0.15){\circle*{0.1}}
			\put(2,-1.30){\circle*{0.1}}
			\put(2,-2.30){\circle*{0.1}}
			\put(2,-3.30){\circle*{0.1}}
			%
			\put(-1.5,3){\circle*{0.1}}
			\put(-1.5,2){\circle*{0.1}}
			\put(-1.5,1){\circle*{0.1}}
			\put(-1.5,-0.15){\circle*{0.1}}
			\put(-1.5,-1.30){\circle*{0.1}}
			\put(-1.5,-2.30){\circle*{0.1}}
			\put(-1.5,-3.30){\circle*{0.1}}
			%
			%
			%
		\end{pspicture}
		\vspace{0.4cm}
		\caption{Poles in the $\omega$-complex plane for the $\kappa$-retarded Green function with $|{\bf k}|<(2q)^{-1}$.}
		\label{contornoporbaixo}
	\end{center}
\end{figure}
The poles of the $\kappa$-retarded Green function are given by
\begin{subequations}
	\begin{eqnarray}
		\omega_{n}^{(\pm)} \!&=&\! \omega^{(\pm)}({\bf k})+\frac{i\pi n}{q} -i\varepsilon \; , \; \mbox{if} \; |{\bf k}|<(2q)^{-1} \; ,
		\label{omeganpmret}
		\\
		\omega_{n} \!&=&\! -\frac{1}{2q}\ln(2)+\frac{i\pi n}{q}-i\varepsilon \; , \; \mbox{if} \; |{\bf k}|=(2q)^{-1} \; ,
		\label{omeganret}
		\\
		\omega_{n}^{(>)} \!&=&\! \omega^{(>)}({\bf k})-\frac{i\pi}{2q}+\frac{i\pi n}{q}-i\varepsilon \; , \; \mbox{if} \; |{\bf k}|=(2q)^{-1} \; .
		\label{omegan>ret}
	\end{eqnarray}
\end{subequations}
%

%
%
%
\begin{figure}[!h]
	\begin{center}
		\newpsobject{showgrid}{psgrid}{subgriddiv=1,griddots=10,gridlabels=6pt}
		\begin{pspicture}(-2,-3)(4,3.3)
			\psset{arrowsize=0.2 2}
			%
			%
			\psline[linecolor=black,linewidth=0.4mm]{->}(0.5,-3.5)(0.5,3.4)
			\psline[linecolor=black,linewidth=0.4mm]{->}(-4,0)(5,0)
			%
			%
			\psline[linecolor=black,linewidth=0.5mm](-1.5,0)(-1.5,-0.15)
			%
			%
			\put(5,-0.3){$\Re(\omega)$}
			\put(0.7,3.1){$\Im(\omega)$}
			\put(3,1){$(2q)^{-1}<|{\bf k}|<q^{-1}$}
			\put(2.2,-0.7){$\omega^{(>)}({\bf k})-\frac{i\pi}{2q} -i\varepsilon$}
			\put(-3.6,-0.5){$\omega^{(-)}({\bf k})-i\varepsilon$}
			%
			\put(2,2.65){\circle*{0.1}}
			\put(2,1.65){\circle*{0.1}}
			\put(2,0.65){\circle*{0.1}}
			\put(2,-0.65){\circle*{0.1}}
			\put(2,-1.65){\circle*{0.1}}
			\put(2,-2.65){\circle*{0.1}}
			%
			\put(-1.5,3){\circle*{0.1}}
			\put(-1.5,2){\circle*{0.1}}
			\put(-1.5,1){\circle*{0.1}}
			\put(-1.5,-0.15){\circle*{0.1}}
			\put(-1.5,-1.30){\circle*{0.1}}
			\put(-1.5,-2.30){\circle*{0.1}}
			\put(-1.5,-3.30){\circle*{0.1}}
			%
			%
			%
		\end{pspicture}
		\vspace{0.4cm}
		\caption{Poles in the $\omega$-complex plane for the $\kappa$-retarded Green function with $(2q)^{-1}<|{\bf k}|<q^{-1}$.}
	\end{center}
\end{figure}
The structure of poles on the complex plane of $\omega$,
for the cases $|{\bf k}|=(2q)^{-1}$, $|{\bf k}|<(2q)^{-1}$ and $|{\bf k}|>(2q)^{-1}$
is showed in the figures below, respectively. Except the first figure, the others
one show the asymmetric position of the poles on the $\omega$-real axis. 
When $|{\bf k}| \rightarrow (2q)^{-1}$, the poles of positive frequencies go away from imaginary axis, while the
poles of negative frequencies go to $-(2q)^{-1}\ln(2)$. Analogously, we have the figure of poles for $|{\bf k}|>(2q)^{-1}$.
In this case, we observe a sign exchange, in which $\omega^{(>)}$ is positive
for $(2q)^{-1}<|{\bf k}|<q^{-1}$, as is illustrated above, and it becomes
negative for $|{\bf k}|>q^{-1}$. By the last conclusions, we will make firstly the $\omega$-integration of (\ref{GIntomega}) in the case of $|{\bf k}|=(2q)^{-1}$, 
and after that, in the two ranges $|{\bf k}|<(2q)^{-1}$ and $|{\bf k}|>(2q)^{-1}$. The cases are discussed below :

\begin{enumerate}
	\item When $|{\bf k}|=(2q)^{-1}$ is a special case. In this case, the original Green equation (\ref{EqGreenBDPCk=kappa})
	is just a temporal equation, and the $\nabla^2$-operator is substituted by $(2q)^{-1}$. Thereby, we have
	\begin{equation}\label{EqGreenBDPCk=kappa}
		\left[\, \frac{1}{q^2} \, \sin^2\left(q \, \frac{\partial}{\partial t}\right)-(2q)^{-2} \, e^{i2q\partial_{t}} \, \right]
		G_{|{\bf k}|=(2q)^{-1}}(t-t^{\prime})=- \, \delta(t-t^{\prime}) \; ,
	\end{equation}
	and we use the Fourier transform to obtain the $\omega$-integration (\ref{GIntomega})
	with $|{\bf k}|=(2q)^{-1}$, {\it i.e.}, we make $1-4q^{2}\,{\bf k}^2=0$ in the denominator
	of (\ref{denominator}), so the $\omega$-integration is
	\begin{equation}\label{GIntomegak=kappa}
		{\cal G}^{(+)}[(2q)^{-1},t-t^{\prime}]=-2q^{2}\int_{(+)}
		\frac{d\omega}{2\pi} \; e^{-i\omega  (t-t^{\prime})} \, \frac{e^{2q\omega}}{e^{2q\omega}-e^{-\ln(2)-i2q\varepsilon}} \; .
	\end{equation}
	
	By convenience, we make transformation $\xi =2q\omega$, that permit us
	to write (\ref{GIntomegak=kappa}) as
	\begin{equation}\label{GIntomegak=kappaxi}
		{\cal G}^{(+)}[(2q)^{-1},t-t^{\prime}]=-q\int_{(+)}
		\frac{d\xi}{2\pi} \; e^{-i \frac{\xi}{2q}  (t-t^{\prime})}
		\frac{d}{d\xi} \ln\left(e^{\xi}-e^{-\ln(2)-i2q\varepsilon}\right) \; .
	\end{equation}
	
	This integral is in the form for the application of Cauchy's integral, that has the following
	formula
	\begin{equation}\label{Cauchy1}
		\oint_{\cal C}\frac{dz}{2\pi}f(z)\frac{d}{dz}\ln\left[\phi(z)\right]=-i\sum_{n}f(p_n)+i\sum_{n}f(z_n) \; ,
	\end{equation}
	%
	%
	%
	%
	\begin{figure}[!h]
		\begin{center}
			\newpsobject{showgrid}{psgrid}{subgriddiv=1,griddots=10,gridlabels=6pt}
			\begin{pspicture}(-2,-3)(4,3.3)
				\psset{arrowsize=0.2 2}
				%
				%
				\psline[linecolor=black,linewidth=0.4mm]{->}(3,-3.5)(3,3.4)
				\psline[linecolor=black,linewidth=0.4mm]{->}(-4,0)(5,0)
				%
				%
				\psline[linecolor=black,linewidth=0.4mm](-1.5,0)(-1.5,-0.15)
				%
				%
				\put(5,-0.5){$\Re(\omega)$}
				\put(3.2,3.1){$\Im(\omega)$}
				\put(1,1){$|{\bf k}|>q^{-1}$}
				\put(0.7,-0.7){$\omega^{(>)}({\bf k})-\frac{i\pi}{2q}-i\varepsilon$}
				\put(-3.6,-0.5){$\omega^{(-)}({\bf k})-i\varepsilon$}
				%
				\put(0.5,2.65){\circle*{0.1}}
				\put(0.5,1.65){\circle*{0.1}}
				\put(0.5,0.65){\circle*{0.1}}
				\put(0.5,-0.65){\circle*{0.1}}
				\put(0.5,-1.65){\circle*{0.1}}
				\put(0.5,-2.65){\circle*{0.1}}
				%
				\put(-1.5,3){\circle*{0.1}}
				\put(-1.5,2){\circle*{0.1}}
				\put(-1.5,1){\circle*{0.1}}
				\put(-1.5,-0.15){\circle*{0.1}}
				\put(-1.5,-1.30){\circle*{0.1}}
				\put(-1.5,-2.30){\circle*{0.1}}
				\put(-1.5,-3.30){\circle*{0.1}}
				%
				%
			\end{pspicture}
			\vspace{0.5cm}
			\caption{Poles in the $\omega$-complex plane with $|{\bf k}|>q^{-1}$.
				In this case, the poles associated with the positive frequencies $\omega^{(>)}$ cross the imaginary axis, and they are negative.}
		\end{center}
	\end{figure}
	where $f$ is the analytic function inside and on the ${\cal C}$-contour, $\phi$ is analytic function
	inside and on the ${\cal C}$-contour except at a finite number of poles. The summations are on the poles $(p_n)$, 
	and zeros $(z_n)$ of the function $\phi$ inside the ${\cal C}$-contour. In this simplest case, we sum just
	the zeros that are exactly the poles (\ref{omeganret}) multiplied by $2q$, {\it i.e.}, $\xi_{n}=-\ln(2)+i2\pi n-i2q \varepsilon$, 
	with $n$ any integer. In our calculation, we increase the contour ${\cal C}$
	in (\ref{Cauchy1}) tends to a convergent series. For $t-t^{\prime}>0$, the integrations
	on the real axis of (\ref{GIntomegak=kappaxi}) are extended to a closed contour ${\cal C}_{1}$ with the usual infinite radius 
	semicircle on the lower imaginary half-plane, thereby circling the zeros
	given by $n=\left\{0,-1,-2,...\right\}$. Therefore the theorem
	(\ref{Cauchy1}) applied in the contour ${\cal C}_{1}$ of the figure (\ref{contornocima}), leads to the result
	\begin{equation}\label{Gt-t>0}
		\int_{-\infty}^{\infty} \frac{d\xi}{2\pi} \; e^{-i \xi \frac{(t-t^{\prime})}{2q}} \, \frac{d}{d\xi} \ln\left[\,e^{\xi}-e^{-\ln(2)-i2q\varepsilon}\,\right]=
		-\frac{i\,e^{\frac{i}{2q} \, \ln(2) \, (t-t^{\prime})}}{1-e^{-\frac{\pi(t-t^{\prime})}{q}}} \; , \; \mbox{if} \; \, t-t^{\prime}>0 \; .
	\end{equation}
	%
	%
	%
	%
	\begin{figure}[!h]
		\begin{center}
			\newpsobject{showgrid}{psgrid}{subgriddiv=1,griddots=10,gridlabels=6pt}
			\begin{pspicture}(-2,-3)(4,4)
				\psset{arrowsize=0.2 2}
				%
				\psline[linecolor=black,linewidth=0.4mm]{->}(1,-3.5)(1,4)
				\psline[linecolor=black,linewidth=0.4mm]{->}(-5,0)(6,0)
				%
				\psline[linecolor=black,linewidth=0.4mm]{->}(1,0)(2.6,-0.42)
				\psline[linecolor=black,linewidth=0.4mm](2.5,-0.4)(3.9,-0.8)
				\psline[linecolor=black,linewidth=0.4mm](1,0)(-0.5,-0.4)
				\psline[linecolor=black,linewidth=0.4mm]{<-}(-0.3,-0.35)(-1.9,-0.8)
				\psline[linecolor=black,linewidth=0.5mm](-0.51,0)(-0.5,-0.15)
				%
				\psarc[linecolor=black,linewidth=0.4mm](1,0){3}{195}{270}
				\psarc[linecolor=black,linewidth=0.4mm]{<-}(1,0){3}{270}{345}
				\psarc[linecolor=black,linewidth=0.4mm]{<-}(1,0){2.5}{345}{360}
				\psarc[linecolor=black,linewidth=0.4mm]{->}(1,0){2.5}{180}{195}
				\put(5.5,-0.5){$\Re(\omega)$}
				\put(1.2,3.6){$\Im(\omega)$}
				\put(3.8,-0.5){$\theta_{0}$}
				\put(-2.05,-0.5){$\theta_{0}$}
				%
				\put(2.7,-3){${\cal C}_{1}$}
				\put(-1.8,0.33){$-\kappa\ln(2)-i\varepsilon$}
				%
				\put(-0.5,2.85){\circle*{0.1}}
				\put(-0.5,1.85){\circle*{0.1}}
				\put(-0.5,0.85){\circle*{0.1}}
				\put(-0.5,-0.15){\circle*{0.1}}
				\put(-0.5,-1.15){\circle*{0.1}}
				\put(-0.5,-2.15){\circle*{0.1}}
				\put(-0.5,-3.15){\circle*{0.1}}
				%
				%
				\put(3.8,-2){$t-t^{\prime}>0$}
			\end{pspicture}
			\vspace{0.2cm}
			\caption{Contour in the $\omega$-complex plane, when $t-t^{\prime}>0$.}
			\label{contornocima}
		\end{center}
	\end{figure}
	For $t-t^{\prime}<0$, we use a closed contour with the radius semicircle ${\cal C}_{2}$ on the upper half-plane circling
	the zeros in $n=1,2,...$. Applying Cauchy's integral (\ref{Cauchy1}) to those
	contour integrals, we obtain
	%
	%
	%
	%
	\begin{figure}[!h]
		\begin{center}
			\newpsobject{showgrid}{psgrid}{subgriddiv=1,griddots=10,gridlabels=6pt}
			\begin{pspicture}(-2,-3)(4,4.0)
				\psset{arrowsize=0.2 2}
				%
				%
				\psline[linecolor=black,linewidth=0.4mm]{->}(1,-3)(1,4)
				\psline[linecolor=black,linewidth=0.4mm]{->}(-5,0)(6,0)
				%
				\psline[linecolor=black,linewidth=0.4mm]{->}(1,0)(2.6,0.42)
				\psline[linecolor=black,linewidth=0.4mm](2.5,0.4)(3.9,0.8)
				\psline[linecolor=black,linewidth=0.4mm](1,0)(-0.5,0.4)
				\psline[linecolor=black,linewidth=0.4mm]{<-}(-0.3,0.35)(-1.9,0.8)
				\psline[linecolor=black,linewidth=0.5mm](-0.51,0)(-0.5,-0.15)
				%
				\psarc[linecolor=black,linewidth=0.4mm]{->}(1,0){3}{15}{90}
				\psarc[linecolor=black,linewidth=0.4mm](1,0){3}{90}{165}
				\psarc[linecolor=black,linewidth=0.4mm]{->}(1,0){2.5}{0}{15}
				\psarc[linecolor=black,linewidth=0.4mm]{<-}(1,0){2.5}{165}{180}
				\put(5.5,-0.5){$\Re(\omega)$}
				\put(1.2,3.6){$\Im(\omega)$}
				\put(3.8,0.2){$\theta_{0}$}
				\put(-2.05,0.2){$\theta_{0}$}
				\put(2.3,3.1){${\cal C}_{2}$}
				\put(-1.8,-0.7){$-\kappa\ln(2)-i\varepsilon$}
				\put(-0.5,2.85){\circle*{0.1}}
				\put(-0.5,1.85){\circle*{0.1}}
				\put(-0.5,0.85){\circle*{0.1}}
				\put(-0.5,-0.15){\circle*{0.1}}
				\put(-0.5,-1.15){\circle*{0.1}}
				\put(-0.5,-2.15){\circle*{0.1}}
				\put(-0.5,-3.15){\circle*{0.1}}
				%
				%
				\put(3.5,2){$t-t^{\prime}<0$}
			\end{pspicture}
			\vspace{0.2cm}
			\caption{Contour in the $\omega$-complex plane when $t-t^{\prime}<0$.}
		\end{center}
	\end{figure}
	\begin{equation}\label{Gt-t<0}
		\int_{-\infty}^{\infty}
		\frac{d\xi}{2\pi} \; e^{-i \xi \frac{(t-t^{\prime})}{2q}}
		\frac{d}{d\xi} \ln\left[ \, e^{\xi}-e^{-\ln(2)-i2q\varepsilon} \, \right]=
		i\,e^{\frac{i}{2q}\,\ln(2)(t-t^{\prime})}\,\frac{1}{2}\left[\,\coth\left(\frac{\pi(t^{\prime}-t)}{2q}\right)-1\, \right]  \; .
	\end{equation}
	
	The general solution is the combination of (\ref{Gt-t>0}) and (\ref{Gt-t<0}). The result is
	\begin{equation}\label{GIntomegak=kapparesult}
		{\cal G}^{(+)}[\,(2q)^{-1},t-t^{\prime}\,]=iq \, e^{\frac{i}{2q}\,\ln(2) (t-t^{\prime})} \, \Theta_{q}(t-t^{\prime}) \; ,
	\end{equation}
	where we have defined $\Theta_{q}(t-t^{\prime})$ as the combination of Heaviside functions 
	\begin{equation}\label{Thetaq}
		\Theta_{q}(t-t^{\prime}):=
		\frac{\Theta(t-t ^{\prime})}{1-\mbox{e}^{-\frac{\pi (t-t ^{\prime})}{q}}}
		+\Theta(t^{\prime}-t )\,\frac{1}{2}
		\left[\,\coth\left(\frac{\pi(t ^{\prime}-t )}{2q}\right)-1\,\right] \; .
	\end{equation}
	
	In the limit $q \rightarrow 0$, we recover $\lim_{q\rightarrow 0} \Theta_{q}(t-t^{\prime})=\Theta(t-t^{\prime})$, and the contribution of $t^{\prime}-t>0$ is null. 
	When $t^{\prime}-t\gg q$, the future contribution of $t^{\prime}-t>0$ in (\ref{GIntomegak=kappa}) decays with the exponential function
	\begin{equation}\label{GIntomegak=kappatq}
		\left. {\cal G}^{(+)}[\,(2q)^{-1},t-t^{\prime}\,] \right|_{t^{\prime}-t\gg q} \sim \Theta(t^{\prime}-t) \, q \, e^{-\pi(t^{\prime}-t)/q} \; .
	\end{equation}
	%
	%
	\item For the case $|{\bf k}|<(2q)^{-1}$, the denominator of (\ref{GIntomega}) can be written as
	\begin{eqnarray}\label{denominatork<kappa}
		&&\frac{1}{q^{-2}\sinh^{2}(q(\omega+i\varepsilon))-e^{2q(\omega+i\varepsilon)}{\bf k}^2}
		=\frac{4q^{2}}{e^{2q\omega^{(+)}({\bf k})}-e^{2q\omega^{(-)}({\bf k})}}\times\notag\\
		&&\times\left[ \, \frac{e^{2q\omega}}{e^{2q\omega}-e^{2q(\omega^{(+)}({\bf k})-i\varepsilon)}}-\frac{e^{2q\omega}}{e^{2q\omega}-e^{2q(\omega^{(-)}({\bf k})-i\varepsilon)}} \, \right]\; ,
	\end{eqnarray}
	and the integral (\ref{GIntomega}) is
	\begin{eqnarray}\label{GIntomegak<kappa}
		&&{\cal G}_{|{\bf k}|<(2q)^{-1}|}^{(+)}({\bf k},t-t^{\prime})=\frac{4q^{2}}{e^{2q\omega^{(+)}({\bf k})}-e^{2q\omega^{(-)}({\bf k})}} \int_{(+)}
		\frac{d\omega}{2\pi} \; e^{-i\omega  (t-t^{\prime})}\times\notag\\
		&&\times\left[ \, \frac{e^{2q\omega}}{e^{2q\omega}-e^{2q(\omega^{(+)}({\bf k})-i\varepsilon)}}-\frac{e^{2q\omega}}{e^{2q\omega}-e^{2q(\omega^{(-)}({\bf k})-i\varepsilon)}} \, \right] \; .
	\end{eqnarray}

	Using the variable transformation, $\xi=2q\omega$, we write (\ref{GIntomegak<kappa}) into the form
	of the Cauchy theorem
	\begin{eqnarray}\label{G+k<1/2q}
		&&
		{\cal G}_{|{\bf k}|<(2q)^{-1}}^{(+)}({\bf k},t-t^{\prime})=
		\frac{e^{-2q\left[\omega^{(+)}({\bf k})+\omega^{(-)}({\bf k})\right]}}{2|{\bf k}|}
		\int_{-\infty}^{\infty} \frac{d\xi}{2\pi} \; e^{-i\frac{\left(t-t ^{\prime}\right)}{2q}\xi} \, \frac{d}{d\xi}
		\ln\left[ \, e^\xi-e^{2q\left(\omega^{(+)}({\bf k})-i\varepsilon\right)} \, \right]
		\nonumber \\
		&&
		-\frac{e^{-2q\left[\omega^{(+)}({\bf k})+\omega^{(-)}({\bf k})\right]}}{2|{\bf k}|}
		\int_{-\infty}^{\infty} \frac{d\xi}{2\pi} \; e^{-i\frac{\left(t -t ^{\prime}\right)}{2q}\xi} \, \frac{d}{d\xi}
		\ln\left[\,e^\xi-e^{2q\left(\omega^{(-)}({\bf k})-i\varepsilon\right)}\, \right] \, .
	\end{eqnarray}
	Applying the Cauchy theorem (\ref{Cauchy1}), the zeros of the function inside the logarithm are 
	$\xi_{n}^{(\pm)}=2q\omega^{(\pm)}({\bf k})+i2\pi n-i2q \varepsilon$. For $t-t^{\prime}>0$, the integrations
	on the real axis for the integrals (\ref{G+k<1/2q}) are extended
	to a closed contour with the usual infinite radius semicircle on the lower imaginary half-
	plane, circling the zeros $\zeta^{(\pm)}_{n}$, with $n=\left\{\,0,-1,-2,...\,\right\}$. For $t-t^{\prime}<0$, we use a closed contour with the radius semicircle 
	on the upper imaginary half-plane circling the zeros $\zeta^{(\pm)}_{n}$, now with $n=1,2,...$. The result is 
	\begin{eqnarray}\label{G+k<1/2qResult}
	&&{\cal G}_{|{\bf k}|<(2q)^{-1}}^{(+)}({\bf k},t-t^{\prime})=-\frac{\mbox{e}^{-2q\left[\omega^{(+)}({\bf k})+\omega^{(-)}({\bf k})\right]}}{|{\bf k}|} \;
	e^{-i\left[\,\frac{\omega^{(+)}({\bf k})+\omega^{(-)}({\bf k})}{2}\,\right]\left(t-t^{\prime}\right)}\times\notag\\
		&&\times\sin\left[ \, \left( \frac{\omega^{(+)}({\bf k})-\omega^{(-)}({\bf k})}{2} \right) |t-t^{\prime}| \, \right]
		\Theta_{q}(t-t^{\prime}) \; .
	\end{eqnarray}
	\item In the case of $|{\bf k}|>(2q)^{-1}$, the denominator of (\ref{GIntomega}) is written as
	\begin{eqnarray}
		&&\frac{1}{q^{-2}\sinh^{2}(q(\omega+i\varepsilon))-e^{2q(\omega+i\varepsilon)}\,{\bf k}^2}=\frac{4q^{2}}{e^{2q\omega^{(>)}({\bf k})}+e^{2q\omega^{(-)}({\bf k})}} \times\notag\\
		&&\times
		\left[\frac{e^{2q\omega}}{e^{2q\omega}-e^{2q(\omega^{(-)}({\bf k})-i\varepsilon)}}-\frac{e^{2q\omega}}{e^{2q\omega}+e^{2q(\omega^{(>)}({\bf k})-i\varepsilon)}}\right]\; ,
	\end{eqnarray}
	in which the integral (\ref{GIntomega}) is
	\begin{eqnarray}\label{GIntomegak>kappa}
		\hspace{-1.5cm}&&{\cal G}_{|{\bf k}|>(2q)^{-1}}^{(+)}({\bf k},t-t^{\prime})=\notag\\
		\hspace{-1.5cm}&&\frac{4q^{2}}{e^{2q\omega^{(>)}({\bf k})}+e^{2q\omega^{(-)}({\bf k})}} \int_{(+)}
		\frac{d\omega}{2\pi} \; e^{-i\omega  (t-t^{\prime})} \left[\frac{e^{2q\omega}}{e^{2q\omega}-e^{2q(\omega^{(-)}({\bf k})-i\varepsilon)}}-\frac{e^{2q\omega}}{e^{2q\omega}+e^{2q(\omega^{(>)}({\bf k})-i\varepsilon)}}\right] \; .
	\end{eqnarray}
	
	Using again that $\xi=2q\omega$, we write (\ref{GIntomegak>kappa}) into the form
	\begin{eqnarray}\label{G+k>1/2q}
		{\cal G}_{|{\bf k}|>(2q)^{-1}}^{(+)}({\bf k},t-t^{\prime})\hspace{-17pt}&&=
		\frac{e^{-2q(\omega^{(>)}({\bf k})+\omega^{(-)}({\bf k}))}}{2|{\bf k}|}
		\int_{-\infty}^{\infty} \frac{d\xi}{2\pi} \; e^{-i\frac{(t -t ^{\prime})}{2q}\xi} \, \frac{d}{d\xi}
		\ln\left[e^\xi-e^{2q\left(\omega^{(-)}({\bf k})-i\varepsilon\right)}\right]
		\nonumber \\
		\hspace{-17pt}&&-
		\frac{e^{-2q(\omega^{(>)}({\bf k})+\omega^{(-)}({\bf k}))}}{2|{\bf k}|}
		\int_{-\infty}^{\infty} \frac{d\xi}{2\pi} \; e^{-i\frac{(t -t ^{\prime})}{2q}\xi} \, \frac{d}{d\xi}
		\ln\left[e^\xi+e^{2q\left(\omega^{(>)}({\bf k})-i\varepsilon\right)}\right] . \hspace{0.8cm}
	\end{eqnarray}
	
	The zeros of these integrals are $\xi_{n}^{(-)}$ and $\xi^{(>)}_{n}=2q\omega^{(>)}({\bf k})-i\pi+i2\pi n-i2q \varepsilon$.
	When $t-t^{\prime}>0$, the integrations on the real axis in (\ref{G+k>1/2q}) are extended
	to a closed contour with the usual infinite radius semicircle on the lower imaginary half-
	plane, thereby circling the zeros $\xi_{n}^{(-)}$ and $\xi^{(>)}_{n}$, with $n=\left\{ \, 0,-1,-2,... \, \right\}$. 
	For $t-t^{\prime}<0$, we use a closed contour with the radius semicircle on the upper imaginary half-plane circling
	the zeros in $n=\left\{ \, 1,2,... \, \right\}$. Using Cauchy's integral (\ref{Cauchy1}) to those
	contour integrals, we obtain the result
	\begin{eqnarray}\label{G+k>1/2qResult}
		&&{\cal G}_{|{\bf k}|>(2q)^{-1}}^{(+)}({\bf k},t-t^{\prime})=\frac{\mbox{e}^{-2q\left[\omega^{(>)}({\bf k})+\omega^{(-)}({\bf k})\right]}}{|{\bf k}|} \,
		e^{-i\left[\frac{\omega^{(>)}({\bf k})+\omega^{(-)}({\bf k})}{2}\right]\left(t-t^{\prime}\right)}\times\notag\\
		&&\times\sin\left[ \, \left( \frac{\omega^{(>)}({\bf k})-\omega^{(-)}({\bf k})}{2}-\frac{i\pi}{4q} \right) |t-t^{\prime}| \, \right]
		\, e^{-\frac{\pi\left(t-t^{\prime}\right)}{4q}} \, \Theta_{q}(t-t^{\prime}) \; ,
	\end{eqnarray}
	where $\Theta_{q}(t-t^{\prime})$ we defined previously in (\ref{Thetaq}). 
	The solution that combines (\ref{G+k<1/2qResult}) and (\ref{G+k>1/2qResult}) is
	\begin{equation}\label{IntegraltemporalIaberta}
		{\cal G}^{(+)}({\bf k},t-t^{\prime})=\Theta\left(\frac{1}{2q}-|{\bf k}|\right)
		{\cal G}_{|{\bf k}|<(2q)^{-1}}^{(+)}({\bf k},t-t^{\prime})
		+\Theta\left(|{\bf k}|-\frac{1}{2q}\right){\cal G}_{|{\bf k}|>(2q)^{-1}}^{(+)}({\bf k},t-t^{\prime}) \; .
	\end{equation}
	Substituting (\ref{IntegraltemporalIaberta}) in (\ref{FourierBDPCInt}), the $\kappa$-retarded Green function is reduced to $k$-integral form
	\begin{equation}\label{GretBDPC}
		G^{(+)}(x-x^{\prime})=- \, 
		\frac{\Theta_{q}(t-t^{\prime})}{4\pi|{\bf x}-{\bf x}^{\prime}|}
		\, \delta_{q}(x-x^{\prime}) \; ,
	\end{equation}
	where we have defined the $\delta_{q}(x-x^{\prime})$-function as 
	\begin{eqnarray}\label{deltaqbbc}
		&&
		\delta_{q}(x-x^{\prime}):=\frac{2}{\pi}\int_{0}^{(2q)^{-1}} dk \,
		\sin\!\left(k|{\bf x}-{\bf x}^{\prime}|\right)(1-4q^2\,k^2)^{1+i(t-t^{\prime})/(4q)}\times
		\nonumber \\
		&&
		\times\sin\left[\, \frac{|t-t^{\prime}|}{2q} \, \tanh^{-1}(2qk) \, \right]-\frac{2}{\pi} \, e^{-\frac{\pi(t-t^{\prime})}{2q}} \int_{(2q)^{-1}}^{\infty} dk\sin\left(k|{\bf x}-{\bf x}^{\prime}|\right) \times\notag\\
		&&\times \left(4q^2\,k^2-1\right)^{1+i(t-t^{\prime})/(4q)}\sin\left[\, \left( \, \frac{1}{2q}\,\coth^{-1}(2qk) +\frac{i\pi}{2q} \, \right)|t-t^{\prime}|\,\right] \; .
	\end{eqnarray}
	
	Notice that (\ref{deltaqbbc}) is reduced to the usual Dirac delta when $q\rightarrow 0$ 
	\begin{equation}\label{deltaq=0}
		\lim_{q \rightarrow 0} \delta_{q}(x-x^{\prime})=\delta\left[t^{\prime}-\left(t-|{\bf x}-{\bf x}^{\prime}|\right)\right] \; .
	\end{equation}
	
	For $q \neq 0$, the integrals (\ref{deltaqbbc}) have the analytical solution
	\begin{equation}
		\delta_{q}(x-x^{\prime})=\frac{2\,R\,\tau}{\pi \, q \left(32q^2-i4\,q\,\tau+\tau^2\right)} \, _2F_1\left(1,3+\frac{i\tau}{4q},3-\frac{i\tau}{4q},\frac{-R^2}{4q^2} \right)
		-\frac{\tau}{q\,R}\, J_{1+i\tau/(4q)}\left( \frac{R}{2q} \right) \; ,
	\end{equation}
where we have defined $R=|{\bf x}-{\bf x}^{\prime}|$ and $\tau=t-t^{\prime}$, $_2F_1$ is a Gauss hypergeometric function, and $J_{\nu}(x)$ is a Bessel function of the first kind with complex order of $\nu=1+i\tau/(4q)$.

	The limit $q\rightarrow 0$ in (\ref{GretBDPC}) recovers the known result of retarded Green function in classical electrodynamics  
	\begin{equation}\label{Gretq=0}
		\left. G^{(+)}(x-x^{\prime})\right|_{q=0}=- \, 
		\frac{\Theta(t-t^{\prime})}{4\pi|{\bf x}-{\bf x}^{\prime}|} \, 
		\delta\left( t^{\prime}-t_{r} \right) \; ,
	\end{equation}
	in which $t_{r}=t-|{\bf x}-{\bf x}^{\prime}|$ is the retarded time for the propagation of a signal with velocity $(c=1)$ between two points in space. 
	The $\kappa$-retarded Green function (\ref{GretBDPC}) exhibits the zero limit for the remote past
	\begin{equation}\label{Gpastlim}
		\lim_{t \rightarrow -\infty} G^{(+)}(x-x^{\prime})=0 \; ,
	\end{equation}
	as in the non-deformed case. 
	The future contribution in the Green function $\kappa$-retarded (\ref{GretBDPC}) dies rapidly
	when the time measurement is much greater than the fundamental length $(q)$, {\it i. e.}, for time-intervals of 
	$t ^{\prime}-t \gg q$, thus we have
	\begin{equation}\label{GretBDPC(t'-t>>q)}
		\left. G^{(+)}(t-t^{\prime},{\bf x}-{\bf x}^{\prime})\right|_{t^{\prime}-t\gg q}
		\sim \Theta(t^{\prime}-t) \;  e^{-\pi(t^{\prime}-t)/q}
		\sqrt{\frac{\pi}{4\,q\,\tau}} \, e^{i\pi/4} \, e^{i(R^2+\tau^2)/(4q\tau)}\, \sin\left( \frac{R}{2q} \right) \, ,
	\end{equation}
	that is the very small contribution from future for the $\kappa$-retarded Green function.
	The light cone of the $\kappa$-retarded Green function (real part) (\ref{GretBDPC}) is illustrated
	in the Fig.(\ref{LightConebbc}). We have the plot of $\Re[\delta_{q}]$ 
	for $({\bf x}^{\prime}={\bf 0},t^{\prime}=0,x^{2}=x^{3}=0$,
	and $x^{1}=x)$, with $q=0.05$ in length unity. As we have expected, the deformation
	smears the light cone in spacetime, and it brings contribution of the future to $\kappa$-retarded
	Green function, when $|{\bf k}|<(2q)^{-1}$ and $|{\bf k}|>(2q)^{-1}$. Since the deformation is
	very small, the second integral in (\ref{deltaqbbc}) is not meaningful in relation to first one.
	%
	%
	%
	\begin{figure}[!hbt]
		\begin{center}
			\includegraphics[width=9.0cm]{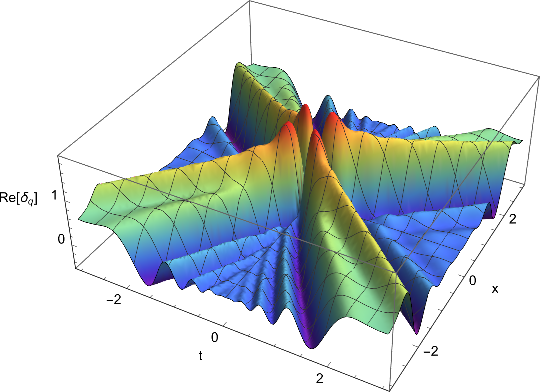}
			\caption{The blurred light cone in the bicrossproduct basis with space $(x)$, time $(t)$ and $q$ in arbitrary units for $x^{\prime}=t^{\prime}=0$.
				A plot of $\Re[\delta_{q}]$ given by (\ref{deltaqbbc}) with $q=0.05$ in length unity.}\label{LightConebbc}
		\end{center}
	\end{figure}
	%
	%
	%
	
\end{enumerate}
The $\kappa$-advanced Green function is calculated in an analogous way, whose the result in terms of quadrature $k$-integral is 
\begin{equation}\label{GadvBDPC}
	G^{(-)}(x-x^{\prime})=-
	\, \frac{\Theta_{q}(t^{\prime}-t)}{4\pi|{\bf x}-{\bf x}^{\prime}|}
	\, \delta_{q}(x-x^{\prime}) \; ,
\end{equation}
where the limit of $q\rightarrow 0$ leads to the known advanced Green function
\begin{equation}\label{FGreenq=0}
	\left. G^{(-)}(x-x^{\prime})\right|_{q=0}=- \, \frac{\Theta(t^{\prime}-t)}{4\pi|{\bf x}-{\bf x}^{\prime}|}
	\, \delta\left(t^{\prime}-t_{a}\right) \; ,
\end{equation}
in which $t_{a}=t+|{\bf x}-{\bf x}^{\prime}|$ is the advanced time. In a distant future, the $\kappa$-advanced Green function (\ref{GadvBDPC})
has the property
\begin{equation}\label{Gfuturelim}
	\lim_{t \rightarrow \infty} G^{(-)}(x-x^{\prime})=0 \; .
\end{equation}

Analogously, for time intervals of $t-t^{\prime} \gg q$, the $\kappa$-advanced Green function has the behavior
\begin{equation}\label{GadvBDPC(t-t'>>q)}
	\left. G^{(-)}(t-t^{\prime},{\bf x}-{\bf x}^{\prime})\right|_{t-t^{\prime}\gg q}
	\sim -\,\Theta(t-t^{\prime}) \;  e^{-\pi(t-t^{\prime})/q} \, 
	\sqrt{\frac{\pi}{4\,q\,\tau}} \, e^{i\pi/4} \, e^{i(R^2+\tau^2)/(4q\tau)}\, \sin\left( \frac{R}{2q} \right) \; ,
\end{equation}
that is the very small contribution from the past.
\section{Conclusions}
\label{sec6}
In this work, we have presented the $\kappa$-deformed electrodynamics equations that leads to dispersion relation in the bicrossproduct basis.
This is related to the $\kappa$-deformed algebra (a Hopf algebra) that is a candidate for describing a new spacetime including
a length scale. The electrodynamics is modified by differential equations of infinite order in the time coordinate, where non-locality 
is introduced by a $q$-parameter with dimension of length, defined as $q=(2\kappa)^{-1}$. We studied aspects of classical $\kappa$-electrodynamics as
plane wave propagation in the vacuum, gauge symmetry, solutions for the potentials and their corresponding Green functions. The plane wave solutions for the electric 
and magnetic fields governed by the equations satisfy the dispersion relation associated with the Casimir operator of the $\kappa$-Poincaré algebra in the bicrossproduct basis. 
The fields remain with transversal property in the vacuum, but the dispersion relation shows that the frequency solutions have a natural cutoff momentum at the scale 
$(2q)^{-1}$ that breaks these solutions. We obtain non-local relations between the electric and magnetic fields, and the scalar and vector potentials, which exhibit a 
$\kappa$-deformed gauge symmetry. Fixing the $\kappa$-Lorenz gauge, the equations for the scalar and vector potentials share the same wave operator associated with the 
original Casimir in the bicrossproduct basis. Afterwards, the frequency solutions yield a refractive index with an imaginary part that show a 
attenuation in the wave amplitudes in which the effect is relevant when the $q$-parameter is of the order of the wavelength. The group and phase velocities 
for the wave exhibit superluminal propagation, with $v_{g}>1$ and $v_{ph}>1$ using natural units $(c=1)$. It is worth noting that this result is typical for non-commutative field theories \cite{NCED3}.                
Then, we calculated the Green functions under the retarded and
advanced prescriptions for the $\kappa$-deformed Casimir operator in the bicrossproduct basis.
These Green functions, which we have called $\kappa$-retarded and $\kappa$-advanced ones,
have the desired limits when the $q$-deformation parameter goes to zero. Moreover, when $q \neq 0$, the deformed Green functions 
exhibit a blurring in the distinction between past and future within time intervals on the order of the
fundamental length $q$-parameter. This result was also confirmed by \cite{Neves1}, for the case of standard basis of the $\kappa$-Poincaré algebra. 
We may say that it is rather surprising and interesting that objects such as these $\kappa$-deformed Green functions, which can be defined from the sole data of a deformed
dispersion relation, exhibit the relaxed causality behavior to be expected in the realm of quantum field theory, although some authors consider such feature as a consequence of promoted to a nonlinear wave operator on a space of commutative functions, where its Green functions are calculated by inverting classical partial differential equations \cite{PLB,NPB,CQG}. For very small deformations, when the $q$-parameter is very small with respect to the time-intervals, the future contribution decays with a exponentially in 
the $\kappa$-retarded Green function, and the past contribution exhibits an analogous exponential decay for the $\kappa$-advanced Green function.      
The future outlook involves to investigate these $\kappa$-Green functions with the formalism of noncommutative geometry, especially those recently developed in \cite{Kurkov2026}. We plan to determine, using the Poisson gauge theory approach \cite{Kupriyanov:2021aet}, whether Poisson field creates emergent gravitational phenomena \cite{Rivelles03,szabo2006} that depends on the $\kappa$-parameter \cite{AN25}. Another perspective is to explore if the confinement property could be obtained from $\kappa$-Minkowski model \cite{Kup2024} and its relation to classical Yang-Mills theories \cite{Schwartz}, as was done for Doplicher-Fredenhagen-Roberts spacetimes \cite{Neves5}, to establish a toy model for strong interactions in other NC spacetimes \cite{AbreuNeves}.
\section*{Acknowledgements}

This work was supported by the Conselho Nacional de Desenvolvimento Cient\'ifico e Tecnol\'ogico (CNPq) under grant 305743/2026-0 (GM) (M. J. Neves).

\end{document}